\documentclass[12pt,preprint]{aastex}

\usepackage{graphicx}
\usepackage{dcolumn}
\usepackage{bm}
\usepackage{ulem}

\begin{document}

\title{Big Bang nucleosynthesis in scalar tensor gravity: the key problem of the primordial $^7$Li abundance}

\author{Julien Larena}
\affil{Laboratoire Univers et Th\'eories, CNRS UMR 8102, Observatoire de
Paris-Meudon, Universit\'e Denis-Diderot Paris 7, 92195 Meudon,
France.}
\email{julien.larena@obspm.fr}
\and
\author{Jean-Michel Alimi}
\affil{Laboratoire Univers et Th\'eories, CNRS UMR 8102, Observatoire de
Paris-Meudon, Universit\'e Denis-Diderot Paris 7, 92195 Meudon,
France.}
\and
\author{Arturo Serna}
\affil{ Departamento de F\'isica y A.C., Universidad Miguel
Hern\'andez, Elche-03202, Spain.
}

\date{\today}

\begin{abstract}
Combined with other CMB experiments, the WMAP survey provides an
accurate estimate of the baryon density of the Universe. In the
framework of the standard Big Bang Nucleosynthesis (BBN), such a
baryon density leads to predictions for the primordial abundances of
$^{4}$He and D in good agreement with observations. However, it also
leads to a significant discrepancy between the predicted and
observed primordial abundance of $^{7}$Li. Such a discrepancy is
often termed as 'the lithium problem'. In this paper, we analyze
this problem in the framework of scalar-tensor theories of gravity.
It is shown that an expansion of the Universe slightly slower than in General
Relativity before BBN, but faster during BBN, solves the lithium
problem and leads to $^4$He and D primordial abundances consistent
with the observational constraints. This kind of behavior is
obtained in numerous scalar-tensor models, both with and without a
self-interaction potential for the scalar field. In models with a
self-interacting scalar field, the convergence towards General
Relativity is ensured without any condition, thanks to an attraction
mechanism which starts to work during the radiation-dominated epoch.
\end{abstract}

\maketitle

\section{Introduction}

The Big Bang Nucleosynthesis prediction for the $^{4}$He primordial
abundance is traditionally considered as a good estimate of the
baryon density of the Universe. However, the recent measurement of
the Cosmic Microwave Background (CMB) anisotropies by WMAP
\citep{WMAPgen} now provides, when combined with other CMB
experiments (CBI and ACBAR), another accurate estimate of the baryon
density: $\Omega_{b}h^{2}=0.0224\pm0.0009$ (or $\eta \times 10^{10}
=6.14\pm 0.25 $ in terms of the baryon to photon ratio)
\citep{WMAPparam}. This value of $\eta$ can be used to compute the
primordial abundances of light elements (mainly $^4$He, $^3$He, D
and $^7$Li). The comparison of predicted and observed abundances is
then a way to test the concordance between BBN and the CMB data.

When one assumes the WMAP estimate of the baryon to photon ratio,
the primordial abundances predicted by the standard BBN ( i.e. three
families of light neutrinos, a neutron mean life $\tau_{n}=885.7\pm
0.8 s$, gravitation described by General Relativity and a
homogeneous and isotropic Universe) are \citep{Cyburt2003}:
\begin{eqnarray}
\mbox{Y}_{p}&=&0.2484^{+0.0004}_{-0.0005} \nonumber \\
\mbox{D/H}&=&2.75^{+0.24}_{-0.19}\times 10^{-5} \nonumber \\
^{7}\mbox{Li/H}&=&3.82^{+0.73}_{-0.60}\times 10^{-10} \nonumber
\end{eqnarray}
while the observed abundances are:
\begin{eqnarray}\label{eq:observed}
\mbox{Y}_{p}&=& \left\{
\begin{array}{ll}
0.2391\pm 0.0020 & \mbox{ ref. \citep{Luridiana2003}}  \\
            0.2452\pm 0.0015 & \mbox{ ref. \citep{Izotov1999}}
\end{array}\right. \nonumber \\
\mbox{D/H}&=&2.78^{+0.44}_{-0.38}\times 10^{-5} \mbox{  ref. \citep{Kirkman2003}} \\
^{7}\mbox{Li/H}&=&\left\{
\begin{array}{ll}
1.23^{+0.68}_{-0.32}\times 10^{-10} & \mbox{ ref. \citep{Ryan2000}}  \\
             2.19^{+0.46}_{-0.38}\times 10^{-10} & \mbox{ ref.
\citep{Bonifacio2002}.}
\end{array}\right.
\nonumber
\end{eqnarray}

Here, the two values for the observed $Y_p$ correspond to
independent estimates in refs. \citep{Luridiana2003} and
\citep{Izotov1999} based on observations of metal-poor extragalactic
ionized hydrogen regions. The D abundance is determined by
observations of remote cosmological clouds on the line of sight of
high redshift quasars. Finally the $^7$Li/H estimate of
\citet{Ryan2000} has been performed by observing halo stars, while
the value of \citet{Bonifacio2002} comes from the observation of
stars in the globular cluster NGC 6397. Although some recent
estimates \citep{Boesgaard2005} leads to a somewhat smaller value,
several other independent determinations
\citep{Ryanal99,Theveninal2001,Asplund2005} obtain observed
primordial abundances of $^7$Li/H similar to those reported either
by \citet{Ryan2000} or  \citet{Bonifacio2002}.

We can see a relatively good agreement for the predicted and
observed abundances of $^{4}$He and D, but a large discrepancy for
$^{7}$Li/H. There had been many attempts to account for the low
$^7$Li abundance indicated by observations. The first and most
conservative possibility is the existence of systematic
uncertainties in the observational determination of the $^7$Li
abundance. However, such uncertainties are not large enough to solve
the problem (see \citet{Ryan2000} and \citet{Bonifacio2002}). In the
same way, the systematic study of uncertainties on the nuclear
reaction rates performed in \citet{Cocal2004} indicates that these
uncertainties cannot explain the large discrepancy emphasized
before. Modifications of the standard nucleosynthesis scenario then
arise as possible ways to reconcile the predictions with the
observations. Inhomogeneous nucleosynthesis has been analyzed
\citep{Jedamzik2001} but overproduces $^7$Li. Late particles decays
can deplete $^7$Li, but they cannot account for the observed D and
D/$^3$He primordial abundances \citep{Jedam2,Jedam3,Ellis2005}; another
possibility is provided by the creation of baryons after BBN, accompanied by a
lepton asymmetry before BBN, the two processes arising from Q-balls \citep{ichikawa}.

In this work, we will address the lithium abundance problem in the
framework of scalar-tensor theories of gravity
\citep{bransdicke61,bergmann68,wagoner70,nordvedt70}. In these
theories, gravitation is modified by the introduction of a scalar
degree of freedom, which does not affect the standard nuclear and
particle physics. Such modifications arise as low-energy limits of
superstrings theories \citep{Schwarz88} and they provide a way to
change the expansion history of the Universe with minimal
assumptions. We will show that such theories contain a mechanism
that could be responsible for a low lithium abundance, despite the
high baryon to photon ratio implied by the WMAP estimate. Big Bang
Nucleosynthesis in the context of a scalar-tensor cosmology has been
extensively studied in the past
\citep{wagoner73,barrow78,sernaalimi1,sernaalimi2,sernaalimi3bis,barrow,Cocal2006},
but the main goal of these works was to constrain the parameters of
the theory thanks to the primordial abundances of $^4$He, D and
$^7$Li, or to obtain the observed abundances with a matter density
of the Universe different from its commonly assumed value. Moreover,
general self-interacting scalar fields were not considered and, as we will
see later, the introduction of such terms in the lagrangian can
provide the mechanism necessary to explain the low $^7$Li abundance.

The paper is arranged as follows. In Section \ref{STmodels}, the
scalar-tensor gravity theories and the implied cosmological models
are presented. Then, in Section \ref{STBBN}, we analyze the lithium
problem in the framework of different kinds of scalar tensor
theories. We show that the solution of this problem requires a non
trivial dynamics for the scalar field at the epoch of BBN. Finally,
we discuss the generality of that solution.

\section{Scalar-tensor cosmological model}\label{STmodels}

\subsection{Equations and observable quantities}
In scalar-tensor theories of gravity, the dynamics of the Universe
contains a new scalar degree of freedom that couples explicitly to
the energy content of the Universe. In units of $c=1$, the action
generically writes, in the Einstein frame:
\begin{eqnarray}
S &=&\frac{1}{4\pi G_{*}}\int \left(\frac{R^{*}}{4}-
\frac{1}{2}\varphi_{,\mu}\varphi^{\mu} -
V(\varphi)\right)\sqrt{-g^{*}}d^{4}x  \nonumber \\ &+&
S_{m}(\psi_{m},A^{2}(\varphi)g_{\mu \nu}^{*}),\label{eq:EFaction}
\end{eqnarray}
$G_{*}$ being a bare gravitational constant, $\varphi$ the scalar
field, $V(\varphi)$ its self-interaction term and $A(\varphi)$ its
coupling to matter. The functional
$S_{m}(\psi_{m},A^{2}(\varphi)g_{\mu \nu}^{*})$ stands for the
action of any field $\psi_{m}$ that contributes to the energy
content of the Universe. It expresses the fact that all these fields
couple universally to a conformal metric $g_{\mu
\nu}=A^{2}(\varphi)g_{\mu \nu}^{*}$, then implying that the weak
equivalence principle (local universality of free fall for
non-gravitationally  bound objects) holds in this class of theories.
The metric $g_{\mu \nu}$ defines the Dicke-Jordan frame, in which
standard rods and clocks can be used to make measurements (since in
this frame, the matter part of the action acquires its standard
form). Defining
\begin{equation}\label{eq:alpha}
\alpha(\varphi)=\frac{d\ln A(\varphi)}{d\varphi},
\end{equation}
and considering the transformations:
\begin{eqnarray}\label{eq:conftrans}
g_{\mu \nu}&=&A^{2}(\varphi)g_{\mu \nu}^{*} \nonumber \\
\phi&=&A^{-2}(\varphi) \nonumber \\
U(\phi)&=&2A^{4}(\varphi)V(\varphi) \nonumber \\
| 3+2\omega(\phi)|&=&\alpha^{-2}(\varphi) \nonumber
\end{eqnarray}
\noindent the action in the Dicke-Jordan frame writes:
\begin{eqnarray}\label{eq:JFaction}
 S&=&\frac{1}{16\pi G_{*}}\int \left(\phi R-\frac{\omega(\phi)}{\phi}\phi_{,\mu}\phi^{\mu}
 -U(\phi)\right)\sqrt{-g}d^{4}x \nonumber\\
 &+& S_{m}(\psi_{m},g_{\mu \nu})
\end{eqnarray}

Despite the conformal relation, these two frames have a different
status: in the Dicke-Jordan frame, where the gravitational degrees
of freedom are mixed, the lagrangian for the matter fields does not
contain explicitly the new scalar field: the non gravitational
physics has then its standard form. In the Einstein frame, the
scalar degree of freedom explicitly couples to the matter fields,
then leading for example to the variation of the inertial masses of
point-like particles. Of course, the two frames describe the same
physical world. Nevertheless, the usual interpretation of the
observable quantities is profoundly modified in the Einstein frame,
whereas it holds in the Dicke-Jordan frame, where the rods and
clocks made with matter are not affected by the presence of the
scalar field. That is why we will refer to the Dicke-Jordan frame as
the observable one.

Varying the Einstein frame action (\ref{eq:EFaction}) with respect
to the fields yields the equations:
\begin{eqnarray}\label{eq:fieldseq}
R^{*}_{\mu\nu}-\frac{1}{2}R^{*}g^{*}_{\mu\nu} &=& 8\pi G_{*}T^{*}_{\mu\nu}+T^{\varphi}_{\mu \nu} \\
\Box^{*}\varphi&=&-4\pi G_{*}\alpha(\varphi)T_{*}+\frac{dV(\varphi)}{d\varphi} \\
\nabla^{*}_{\nu}T^{\nu}_{*\mu}&=&\alpha(\varphi)T_{*}\nabla^{*}_{\mu}\varphi
\end{eqnarray}
where $T_{*}$ is the trace of the energy-momentum tensor of matter
fields $T_{*\mu\nu}$, related to the observable one by
$T_{\mu\nu}=A^{-2}(\varphi)T_{*\mu\nu}$, and $T^{\varphi}_{\mu
\nu}=2\varphi_{,\mu}\varphi_{,\nu}-g^{*}_{\mu\nu}(g_{*}^{\alpha
\beta}\varphi_{\alpha}\varphi_{\beta})-2V(\varphi)g^{*}_{\mu\nu}$ is
the energy-momentum tensor of the scalar field. It is important to
note that these equations reduces to those of General Relativity in
presence of a scalar field if $\alpha(\varphi)=0$.

Except when the contrary is explicitly stated, all the computations
presented in this paper have been performed in the Einstein frame,
because it leads to well posed Cauchy problems (that is elliptic
and/or hyperbolic equations with a set of initial conditions) and
has a perfectly regular dynamics. Nevertheless, the cosmological
evolution resulting from these computations was later expressed in
the Dicke-Jordan frame, where the interpretation of the observable
quantities is easier.

\subsection{Homogeneous and isotropic Universe}
Under the assumption of a Universe filled with various homogeneous
and isotropic matter fluids, the metric reduces to the
Friedman-Lema\^{\i}tre-Robertson-Walker (FLRW) form in the
observable Dicke-Jordan frame:
\begin{eqnarray}
ds^{2}&=&-dt^{2}+a^{2}(t)dl^{2} \nonumber \\
dl^{2}&=&\frac{dr^{2}}{1-kr^{2}}+r^{2}(d\theta^{2}+sin^{2}\theta d\psi^{2}) \nonumber
\end{eqnarray}
and, also in the Einstein frame because of the conformal relation:
\begin{eqnarray*}
ds_{*}^{2}&=&-dt_{*}^{2}+a_{*}^{2}(t*)dl^{2}
\end{eqnarray*}
with $dt=A(\varphi(t_{*}))dt_{*}$ and
$a(t)=A(\varphi(t_{*}))a_{*}(t_{*})$.

Then, defining $H_{*}=(1/a_{*})(da_{*}/dt_{*})$, the fields
equations (\ref{eq:fieldseq}) become:
\begin{eqnarray}
H_{*}^{2}=\frac{8\pi G_{*}\rho_{*}}{3}+\frac{1}{3}\dot{\varphi}^{2} + \frac{2}{3}V(\varphi)-\frac{k}{a_{*}^{2}(t_{*})} \label{eq:cosmoeq} \\
-3\frac{\ddot{a_{*}}}{a_{*}}=4\pi G_{*}(\rho_{*}+3p_{*})+2\dot{\varphi}^{2}-2V(\varphi) \\
\ddot{\varphi}+3H_{*}\dot{\varphi}+\frac{dV(\varphi)}{d\varphi}=-4\pi
G_{*}\alpha(\varphi)(\rho_{*}-3p_{*})\label{eq:evequations}
\end{eqnarray}
where dots denote derivatives with respect to $t_{*}$, whereas
$\rho_{*}$ and $p_{*}$ are the total mass-energy density and
pressure, respectively. They are conformally related to the energy
density $\rho$ and the pressure $p$ in the Jordan frame by
$\rho_{*}=A^{4}(\varphi)\rho$ and $p_{*}=A^{4}(\varphi)p$. The
observable Hubble parameter $H(t)$ is related to the Einstein frame
quantities by
\begin{equation}\label{eq:obshubble}
H(t)=\frac{1}{A(\varphi)}(H_{*}(t_{*})+\alpha(\varphi(t_{*}))\dot{\varphi}(t_{*}))
\end{equation}

\subsection{Dynamics of the scalar field}\label{sec:stdyn}

In the form given above, the time evolution of the scalar field is
coupled, both in the Dicke-Jordan and the Einstein frame, to that of the
scale factor. Previous works \citep{sernaalimi3bis,damournordtvedt93}
have found that, by introducing an appropriate change of variables,
it is possible to obtain an evolution equation for the scalar field
which is independent of the cosmic scale factor. Such an equation
allows for the dynamical analysis of any scalar tensor theory and,
in particular, its asymptotic behavior at early and late times. When
the latter behavior implies a convergence mechanism towards General
Relativity, one can ensures that the Solar System constraints will
be satisfied. We will now extend such previous works to the general
case of scalar-tensor theories with a self-interaction term.

Introducing a new evolution parameter $\lambda=\ln a_{*}+const$, so
that $d\lambda=H_{*}dt_{*}$, and denoting $u'=du/d\lambda$, the
evolution equations (\ref{eq:cosmoeq})-(\ref{eq:evequations}) lead
to:
\begin{eqnarray}\label{eq:dynscal}
\frac{2(1-\epsilon+\eta)}{3-\varphi '^{2}}\varphi''&+(1-w_{b}-\frac{4}{3}\epsilon+2\eta)\varphi'& \nonumber \\
 =&-\Theta-(1-3w_{b})\alpha(\varphi)&
\end{eqnarray}
where
\begin{eqnarray*}
\epsilon(\lambda)&=&\frac{3k}{8\pi G_{*}\rho_{*}(\lambda)a_{*}^{2}(\lambda)} \\
\eta(\varphi,\lambda)&=&\frac{V(\varphi)}{4\pi G_{*}\rho_{*}(\lambda)}\\
w_{b}&=&\frac{p_{*}}{\rho_{*}}\\
\Theta(\varphi,\lambda)&=&\frac{\frac{dV(\varphi)}{d\varphi}}{4\pi G_{*}\rho_{*}(\lambda)}
\end{eqnarray*}

This equation is similar, but not equivalent, to the motion equation
of a mechanical oscillator with a varying "mass"
$m_{eff}=2(1-\epsilon+\eta)/(3-\varphi'^{2})$, a varying "friction"
$\nu_{eff}=(1-w_{b}-4\epsilon/3+2\eta)$ and a "force" term
$F_{eff}=-\Theta-(1-3w_{b})\alpha$. Then, one can define an
effective potential $V_{eff}(\varphi,\lambda)$ that verifies the
relation:
\begin{eqnarray} \label{eq:effpot}
\frac{\partial V_{eff}(\varphi,\lambda)}{\partial \varphi}&=&\Theta(\varphi,\lambda)+(1-3w_{b})\alpha(\varphi)
\end{eqnarray}

In the following, we will only consider flat cosmologies
($\epsilon=0$). In these cases, the "friction" term
$(1-w_{b}+2\eta)$ is always positive, and the dynamics of the scalar
field is then analogous to a damped oscillating motion in the
effective potential $V_{eff}(\varphi,\lambda)$. Such an effective
potential (see Eq. (\ref{eq:effpot})) presents two different parts: a
term due to the coupling function $\alpha(\varphi)$ and another term
due to the self-interaction $V(\varphi)$ of the scalar field.

During the matter-dominated era ($w_b=0$), both terms are important
to determine the effective potential. The minima of $V_{eff}$ will
be attractors for the dynamics of the scalar field and determine the
behavior of the theory at late times. Consequently, if the
relativistic value $\varphi=0$ is a minimum of this effective
potential, the theory will converge towards General Relativity.

Instead, during the radiation-dominated era ($w_b=1/3$), the only
non-vanishing contribution to the effective potential is that due
the self-interaction of the scalar field. Equation
(\ref{eq:dynscal}) then reduces to:
\begin{equation}\label{eq:earlydyn}
\frac{2(1+\eta)}{3-\varphi'^2}\varphi''+\frac{2}{3}(1+3\eta)\varphi'=-\Theta
\end{equation}

At very early times, when $\rho_*$ is very high,
$\Theta\sim\eta\sim0$ so that self-interaction has a negligible
effect on the scalar field dynamics. In this case
($\Theta\sim\eta\sim0$), the integration of equation
(\ref{eq:earlydyn}) gives:
\begin{equation}\label{eq:varphip}
\varphi'^2=\frac{3k^2}{e^{2\lambda}+k^2}
\end{equation}
where $k$ is a constant related to the value of $\varphi'$ at
$\lambda=0$ through $k^2=\varphi'^{2}_{0}/(3-\varphi'^{2}_{0})$.
Note that the above equation implies that
$-\sqrt{3}<\varphi'<\sqrt{3}$.

One can see from equation (\ref{eq:varphip}) that the velocity of
the scalar field exponentially decreases with $\lambda$ at very
early times. When the initial ($\lambda=0$) velocity is not very
high (i.e., when it is not close to $\sqrt{3}$), one can assume that
the scalar field reaches the epochs of interest (those prior to BBN
processes) with an almost vanishing velocity:
\begin{equation}\label{eq:initphi}
\varphi'=0 \mbox{$\;\;$(before BBN)}.
\end{equation}

For the sake of simplicity, we will consider throughout this paper a
vanishing initial value of the scalar field velocity (see the
Appendix for some examples where such a condition has been relaxed).
Hereafter, all quantities at this initial time (prior to BBN) will
be expressed with a subscript 'init'.

The condition (\ref{eq:initphi}) does not however imply a constant
$\varphi$ value during the whole radiation-dominated epoch. Since
$\rho_*$ is a decreasing function of $\lambda$, the contribution of
$V(\varphi)$ to the effective potential will become non-negligible
at some 'time' $\lambda$. If $V(\varphi)$ is chosen to be a function
having a minimum at $\varphi_m$, the scalar field will start to be
attracted towards $\varphi_m$. For example, the simple choice of a
power law $V(\varphi)\propto\varphi^{2n}$  for the self-interaction
term, will imply an attraction mechanism towards General Relativity
which starts to work during the radiation-dominated epoch.

\section{BBN in scalar-tensor theories}
\label{STBBN}

We will now analyze the BBN processes, and the resulting primordial
abundances, in the framework of scalar-tensor theories both with and
without a self-interaction term. As stated above, we will consider
throughout this section that the initial value of the scalar field
is a free parameter, while its initial velocity is fixed to
$\dot{\varphi}_{init}=0$. The initial values of the remaining
variables are chosen so that they imply a flat Universe and lead, at
the present temperature $T_{0}=2.725$, to cosmological parameters
given by: $\Omega_{m,0}=0.27$, $\Omega_{\Lambda,0}=0.73$ and $H_0=71
\mbox{ km}\mbox{ s}^{-1}\mbox{Mpc}^{-1}$. We assume an standard
particle content of the Universe, with three families of light
neutrinos, and the WMAP estimate of the present baryon to photon
ratio ($\eta\times10^{10}=6.14\pm0.25$).

The numerical computation consists of two main parts: first, we use
a sixth-order Runge-Kutta integrator to fully determine the
cosmological model and, in particular, the evolution of the
expansion rate. Once determined the cosmological evolution in a
given gravitational theory, we compute the BBN processes and the
resulting primordial abundances of light elements thanks to a
complex network of 28 nuclear reactions, and using the Beaudet and
Yahil scheme \citep{beaudetyahil77}. We used the reaction rates of
\citet{caughlanfowler88} and \citet{kawanomalaney93} and the updated reaction
rates of the NACRE collaboration \citep{NACRE2004}. The only aspect
that differs from the standard BBN in General Relativity is the
expansion rate of the Universe. It impacts on the various nuclear
abundances because they are determined by many reactions whose
efficiency depends on the ratio of the reaction rates $\Gamma _i$
and the expansion rate $H$. Indeed, a reaction with rate $\Gamma _i$
is in equilibrium when $\Gamma_i/H > 1$, whereas the reaction is
frozen when $\Gamma _i/H < 1$ .
This is directly linked with the fact that the expansion dilutes the
particles. Then, a modification in the expansion history may significantly change the nuclear
abundances by modifying the dynamical structure of the network of
reactions, making some reactions more efficient, and limiting
others (it should be noted that this remark is valid
for the nuclear reaction rates, and also for the rates of the weak interaction
processes that take place before BBN and that interconvert neutrons and protons).

In order to characterize the deviation from General Relativity, it
is convenient to introduce the speed-up factor, defined as the ratio
between the expansion rate $H(T)$ and the corresponding expansion
rate in General Relativity $H_{GR}(T)$ at the same temperature:
\begin{equation}
\xi (T)=\frac{H(T)}{H_{GR}(T)}
\end{equation}
When $\xi(T)>1$ ($\xi(T)<1$ ), the Universe expands faster (slower)
than in General Relativity at the temperature $T$.

\subsection{Theories without a self-interaction term}

We will first consider the simplest case of scalar-tensor models
without a self-interaction term. In this case, the effective
potential during the radiation-dominated epoch is (see equation
(\ref{eq:effpot})):
\begin{equation}
V_{eff}(\varphi) = 0
\end{equation}
and, since we are assuming the initial condition (\ref{eq:initphi}),
the scalar field will be frozen to its initial value
$\varphi_{init}$ (except for a slight temporary perturbation during
the annihilation of electrons and positrons around $T=0.1 \mbox{
MeV}$, when $w_{b}$ slightly deviates from $1/3$). Consequently,
$\dot{\varphi}=0$ during all the BBN processes, and the relation
(\ref{eq:obshubble}) reduces to $H=H_{*}/A(\varphi_{i})$.
Introducing the expression (\ref{eq:cosmoeq}) for $H_{*}$ in this
equation, we can write:

\begin{eqnarray}\label{eq:hubCSUF}
H(T)&=&\sqrt{\frac{8\pi G_{*}A^{2}(\varphi_i)}{3}\rho} \\
    &=&A(\varphi_i)H_{GR}(T)\nonumber
\end{eqnarray}
which implies $\xi \sim A(\varphi_i)=$ constant.

\begin{figure}[htbp]
\begin{center}
\includegraphics[width=6cm]{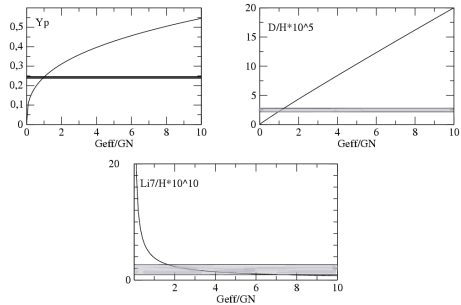}
\end{center}
\caption{Primordial abundances as functions of $G_{eff}/G_{N}$. The
shaded regions are the assumed observed abundances.}
\label{fig:abundvsG}
\end{figure}

Equation (\ref{eq:hubCSUF}) shows that the dynamics of the Universe
is the same as in General Relativity, but with an effective
gravitational constant given by $G_{eff}=G_{*}A^{2}$, where
$G_{*}=G_N$ is the usual Newtonian value ($G_{N}=6.673\times
10^{-11} \mbox{m}^{3}\mbox{ kg}^{-1}\mbox{ s}^{-2}$ ). Cyburt \citep{cyburt2} has determined the scaling of the various
primordial abundances in terms of the physical constants. In
particular, when the dynamics of the Universe is governed by General
Relativity, the scaling with the effective gravitational constant
is:
\begin{eqnarray}\label{eq:Gscale}
\mbox{Y}_{p}&=&0.2484\left(\frac{G_{eff}}{G_{N}}\right)^{0.35}\\
\mbox{D/H}&=&2.75\left(\frac{G_{eff}}{G_{N}}\right)^{0.95}10^{-5}\\
^{3}\mbox{He/H}&=&8.65\left(\frac{G_{eff}}{G_{N}}\right)^{0.34}10^{-6}\\
^{7}\mbox{Li/H}&=&3.82\left(\frac{G_{eff}}{G_{N}}\right)^{-0.72}10^{-10}
\end{eqnarray}
Figure \ref{fig:abundvsG} shows these abundances as functions of
$G_{eff}/G_{N}$, as well as their observational constraints. One
clearly notes from this figure that there is no consistent value for
$G_{eff}/G_N$ that can simultaneously account for all the observed
abundances: the observed $^7$Li primordial abundance requires
$G_{eff}/G_{N}>1.8$ while $^4$He and D impose $G_{eff}/G_{N}\sim 1$.

Consequently, in absence of a self-interaction term, scalar-tensor
theories with an arbitrary coupling function $\alpha(\varphi)$ (with
$\dot{\varphi}_{init}=0$) cannot constitute a solution for the
lithium problem. The same conclusion can be found using the
formulation of Kneller and Steigman \citep{KS2004}

\subsection{Theories with a self-interaction term}\label{sec:selfinteracting}

We will now consider scalar-tensor models with a self-interaction
term. As emphasized above, such a term acts as an effective
potential even during the radiation-dominated epoch. Therefore,
except for very early times (with a very high density), the scalar
field does not remain frozen to its initial value and the speed-up
factor is not necessarily a constant during the BBN processes.

Of course, depending on the functional form of $\alpha(\varphi)$ and
$V(\varphi)$, there exist many theories that exhibit a varying
speed-up factor. Before analyzing a particular choice for
$\alpha(\varphi)$ and $V(\varphi)$, we may ask whether we can
constrain the behavior of the speed-up factor in order to solve the
$^7$Li problem without affecting the other abundances.

\subsubsection{Computation of the primordial abundances}

Since essentially all the neutrons are incorporated into $^4$He
nuclei, the final production of $^4$He is directly related to the
abundance of neutrons in epochs prior to the BBN processes. Indeed,
$Y_p$ is roughly given by
\begin{equation}
Y_p\simeq\frac{2(n/p)_{fr}}{1+(n/p)_{fr}}
\end{equation}
where $(n/p)_{fr}$ denotes the neutron to proton ratio when their
weak reactions, $n+\nu_{e}\leftrightarrow p+e^{-}$ and
$n+e^{+}\leftrightarrow p+\bar{\nu}_{e}$, freeze out of equilibrium.
Such a ratio is then fixed at a temperature $T_{fr}$ of about 1 MeV,
well before the synthesis of light elements (from $\sim100$ KeV to
$\sim 10$ KeV). The $(n/p)_{fr}$ ratio, and hence the final $Y_p$
value, strongly depends on $T_{fr}$ through:
\begin{equation}
(n/p)_{fr}=\exp(-Q/T_{fr})
\end{equation}
where $Q$ is the difference between the neutron and proton masses. A
faster (slower) expansion rate of the Universe implies a higher
(lower) freezing-out temperature, resulting in higher (lower)
neutron and $^4$He abundances.

The $^7$Li production is instead determined by the processes of BBN.
In particular, when the baryon to photon ratio is greater than
$3\times10^{-10}$, the lithium is mainly produced through the
reaction $^3 \mbox{He}(\alpha,\gamma) ^7 \mbox{Be}$ followed by a
decay of $^7$Be through electron captures (when $\eta < 3\times 10^{-10}$, the lithium
is instead produced by $^3 \mbox{He}(\alpha,\gamma))^7\mbox{Li}$).
Therefore, the final abundance of lithium strongly depends on both
the helium production and the efficiency of the $^4$He burning to
give $^7$Li. If $\Gamma_i$ denotes the reaction rate, such an
efficiency is given by $\Gamma_i/H=(\Gamma_i/H_{RG})/\xi(T)$, so
that a faster (slower) expansion rate of the Universe during BBN
implies a less (more) efficient production of $^7$Li.

Since General Relativity leads to a predicted $Y_p$ value in
agreement with observations, one can expect that any other gravity
theory must imply a speed-up factor close to unity at $T\sim 1$ MeV,
or slightly smaller than unity to favor a less efficient production
of $^7$Li without implying an over-production of $^4$He. Later,
during the BBN processes, a speed-up factor that has significantly
increased above unity could help to solve the lithium problem.
Obviously, since any gravity theory must converge at late times
towards General Relativity (in order to be compatible with Solar
System experiments), the increase of the speed-up factor must stop
at some time and, afterwards, it must approach unity. Therefore,
models implying a constant or monotonic $\xi(T)$ function are not
good candidates to solve the lithium problem. We will then restrict
our analysis to self-interacting scalar-tensor theories satisfying
the following two additional conditions: 1) they imply a speed-up
factor with a local maximum (it is a non-monotonic function of $T$),
2) they have an attraction mechanism towards General Relativity.

The analysis of the early behavior of $\xi$ is a complex problem,
specially when $\dot{\varphi}$ starts to deviate from zero (see,
e.g., \citep{sernaalimi2,sernaalimi3} for theories without a
self-interaction term). Nevertheless, the first condition
(non-monotonic $\xi$) is more easily satisfied in theories implying
$\xi<1$ at very early times (when $\dot{\varphi}_{init}\sim 0$ and
the self interaction term is still negligible). Since $\xi\simeq
A(\varphi)$ at such early times, the condition $\xi_{init}<1$
implies $A(\varphi)_{init}<1$. Taking into account equation
(\ref{eq:alpha}), the simple choice of a power law
$\alpha(\varphi)\propto \varphi^{2n}$ for the coupling function,
will imply $\xi_{init}<1$ provided that $\varphi_{init}<0$. On the
other hand, as quoted in section \ref{sec:stdyn}, a similar choice
$V(\varphi)\propto \varphi^{2m}$ for the self interaction term, will
imply the existence during the radiation-dominated epoch of an
attraction mechanism towards General Relativity ($\xi$ will not
remain frozen to its initial value).
Nevertheless, a self-interaction corresponding to $m=1$, in other words, a simple mass
term, is not a viable choice because the scalar field is not damped
efficiently. In fact, by a simple dimensional analysis of equation (\ref{eq:dynscal}), one can define a
characteristic time for the friction:
$\tau_{fric}=\frac{m_{eff}}{\nu_{eff}}$, and a characteristic time for the dragging
in the effective potential:
$\tau_{drag}=\sqrt{\frac{m_{eff}}{|F_{eff}|}}$. When comparing the ratio
$\frac{\tau_{drag}}{\tau_{fric}}$ for a model with $m\geq 2$ and a model with
$m=1$, if the constant coefficients are of the same order of magnitude (which is
required by the fact that the self-interaction must play a role at
temperatures relevant for BBN), one finds, when $\varphi \ll 1$ that:
$$
\left(\frac{\tau_{drag}}{\tau_{fric}}\right)_{m\geq2}>\left(\frac{\tau_{drag}}{\tau_{fric}}\right)_{m=1}
$$
That means that the friction is less efficient in the case $m=1$, when the
scalar field has converged in a neighbourhood of $0$. This is supported by numerical
computations that show that the scalar field quickly reach a neighbourhood of
$0$, but is not damped enough during the radiation dominated era with a
self-interaction of the type $V(\varphi)\propto\varphi^{2}$. In such a case, the
energy contribution of the scalar field becomes more and more important until the beginning of the
matter dominated era. Such models might eventually lead to a very different
estimation of the baryon to photon ratio through CMB anisotropies, then
invalidating the hypothesises at the basis of the computations presented in
this paper.

In order to analyze the possible implications of scalar-tensor
theories with a self-interaction term, we will adopt here the
simple choice:
\begin{eqnarray}\label{eq:theory}
\alpha(\varphi)=a\varphi^2; & V(\varphi)=\Lambda^2\varphi^4
\end{eqnarray}
with $\dot{\varphi}_{init}=0$ and $\varphi_{init}<0$.

The effective potential for these theories writes:
\begin{equation}
V_{eff}(\varphi,\lambda)=(1-3w_b)\frac{a}{3}\varphi^{3}+\frac{\Lambda^{2}
\varphi^{4}}{4\pi G_{*}\rho_{*}(\lambda)},
\end{equation}
which means that, during the radiation dominated period,
$V_{eff}(\varphi,\lambda)=\Lambda^{2} \varphi^{4}/(4\pi
G_{*}\rho_{*})$, leading to an attraction towards $0$ that is
effective as soon as $\Lambda^{2} \varphi^{4} \sim 4\pi
G_{*}\rho_{*}(\lambda)$, or in other words, as soon as the energy
density of the scalar field is of the same order as the energy
density of radiation.

Using (\ref{eq:theory}), and varying the parameters
$(a,\Lambda,\varphi_{init})$, we have numerically computed the BBN
processes and the resulting primordial abundances. We have found
that all models with $-1.5<\varphi_{init}<-0.9$ lead to primordial
abundances that agree with all the observational constraints
mentioned in the Introduction. Consequently, there is not a lithium
problem for such models. The parameters $a$ and $\Lambda$ are not
strongly constrained. The only requirement is that $\Lambda$ should
not be too small in order for the attraction mechanism described
above to occur during BBN, at temperatures between $10$ MeV and $1$
MeV. Indeed, from equation (\ref{eq:dynscal}), we know that the
attraction mechanism approximately (when $|\varphi|\sim1$) occurs
when $\Lambda ^{2}\sim G_{*}\rho_{*}(\lambda)$; for $1\mbox{
MeV}<T<10\mbox{ MeV}$, which correspond to the period of BBN, since
$\rho_{*}(\lambda) \propto (T/T_0)^4$, we are left with
$\sim0.1\mbox{ s}^{-1}<\Lambda<\sim6 \mbox{ s}^{-1}$.

\begin{figure}[htbp]
\begin{center}
\includegraphics[width=6cm,angle=-270]{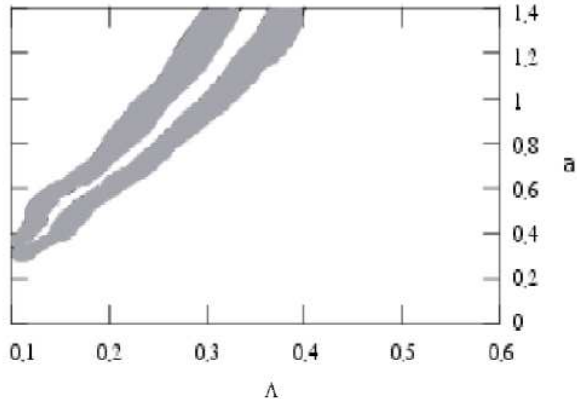}
\end{center}
\caption{Space of the parameters $(a,\Lambda)$ for the theory
defined by $\alpha(\varphi)=a\varphi^2$ and
$V(\varphi)=\Lambda^{2}\varphi^4$ with fixed initial conditions:
$\varphi_{i}=-1.3$ and $\dot{\varphi}_{i}=0$. The acceptable
theories are represented by the shaded regions.}
\label{fig:constraints}
\end{figure}

To be more precise, we now present a special case by fixing the
initial conditions to $\dot{\varphi}_{init}=0$ and
$\varphi_{init}=-1.3$. Figure (\ref{fig:constraints}) displays, in
the $(a,\Lambda)$ plane, the regions of theories leading to the
observed primordial abundances. The constraints only come from the
$^4$He and D abundances, which means that the $^7$Li abundance is in
perfect agreement with the observations for a much wider range of
parameters. The two separated regions correspond to the two
different constraints imposed on the $^4$He abundance
\citep{Luridiana2003,Izotov1999}.

\begin{figure}[htbp]
\begin{center}
\includegraphics[width=8cm,angle=-270]{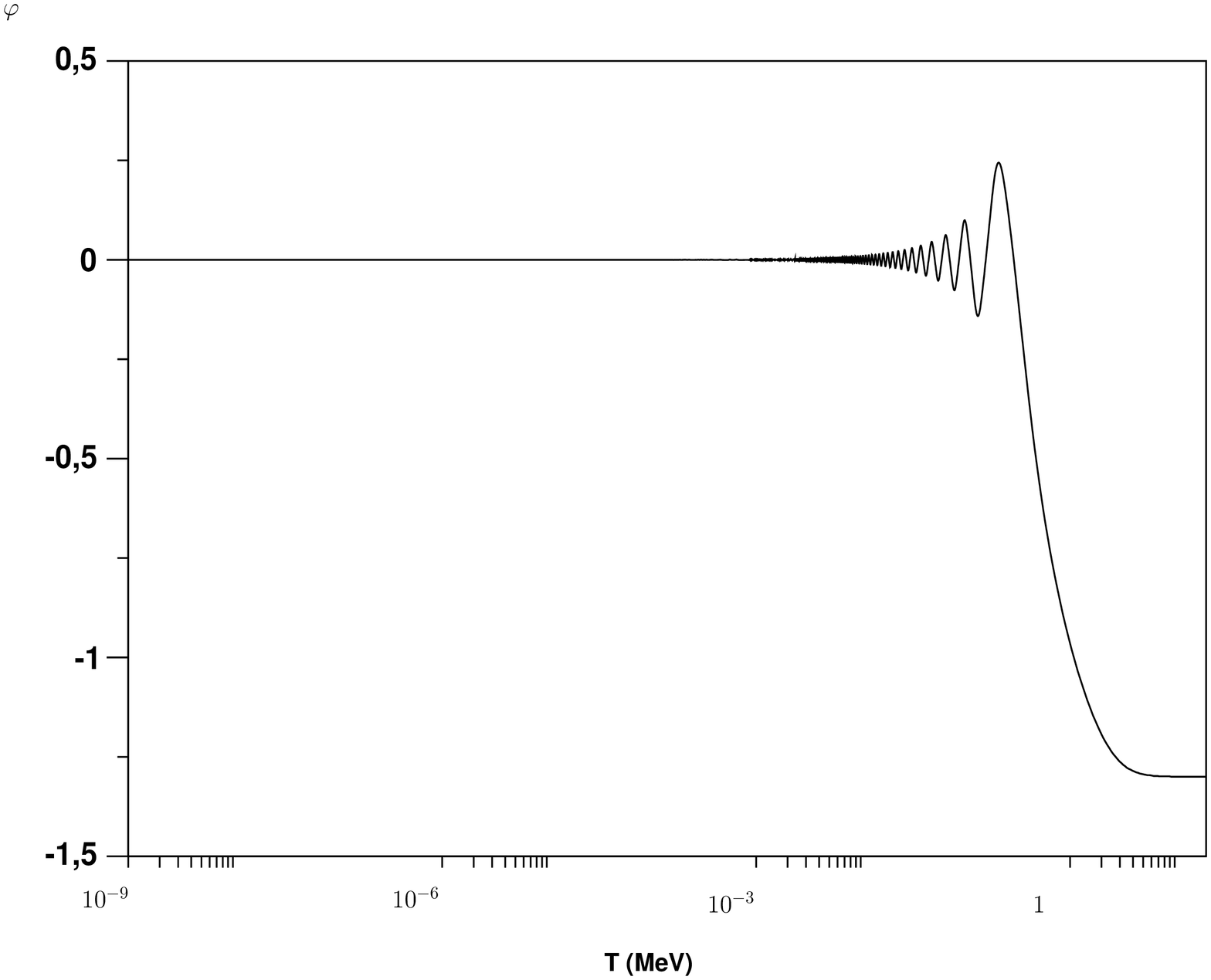}
\end{center}
\caption{Evolution of the scalar field $\varphi$ during BBN for the
scalar tensor theory defined by $\alpha(\varphi)=\varphi^2$ and
$V(\varphi)=0.3^{2}\varphi^4$ with $\varphi_{i}=-1.3$ and
$\dot{\varphi}_{i}=0$.} \label{fig:PhiEF}
\end{figure}

\begin{figure}[htbp]
\begin{center}
\includegraphics[width=8cm,angle=90]{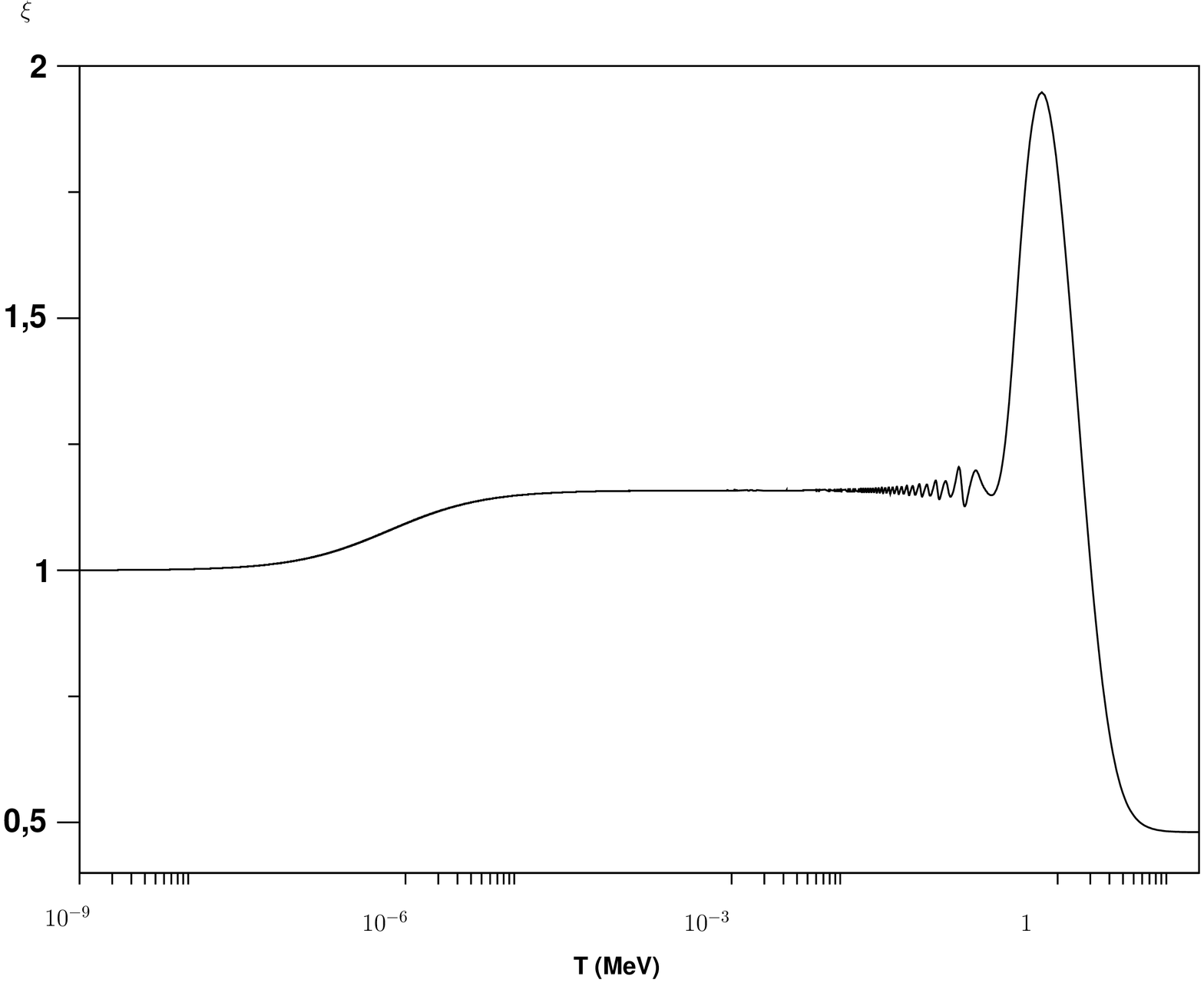}
\end{center}
\caption{Speed-up factor as a function of the temperature for the scalar tensor theory defined by $\alpha(\varphi)=\varphi^2$ and $V(\varphi)=0.3^{2}\varphi^4$ with $\varphi_{i}=-1.3$ and $\dot{\varphi}_{i}=0$. The speed-up factor converges towards $1$ as soon as the matter-radiation equality is reached.}
\label{fig:SUFEF}
\end{figure}

The dynamics of the scalar field and of the speed-up factor are
respectively shown in figures (\ref{fig:PhiEF}) and
(\ref{fig:SUFEF}) for the particular choice
$(a,\Lambda)=(1,0.3\mbox{ s}^{-1})$. In a first phase, the
self-interaction plays no role and the field is almost constant
until $T\sim 5 \mbox{ MeV}$, leading to an expansion slower than in
General Relativity. Then, the $\varphi^4$ term starts to dominate
the effective potential and attracts the scalar field towards zero.
The expansion factor also oscillates and is driven to values greater
than those obtained in General Relativity. The primordial abundances
predicted in this case are $\mbox{Y}_{p}=0.2449$,
$\mbox{D/H}=3.030\times 10^{-5}$ and $^{7}\mbox{Li/H}=2.387\times
10^{-10}$, in agreement with observations.

\begin{figure}[htbp]
\begin{center}
\includegraphics[width=6cm,angle=270]{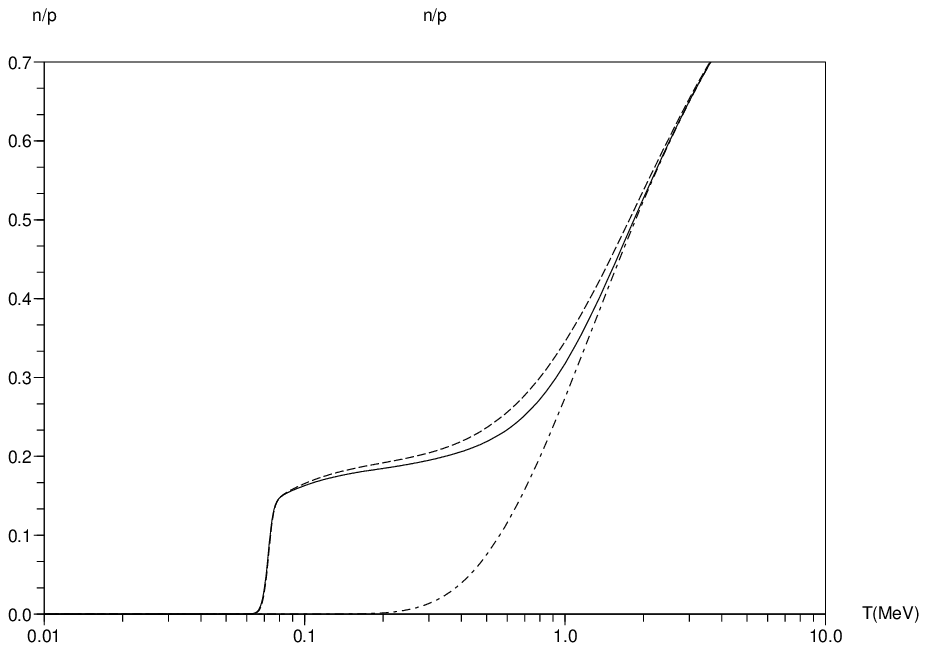}\\
\includegraphics[width=6cm,angle=270]{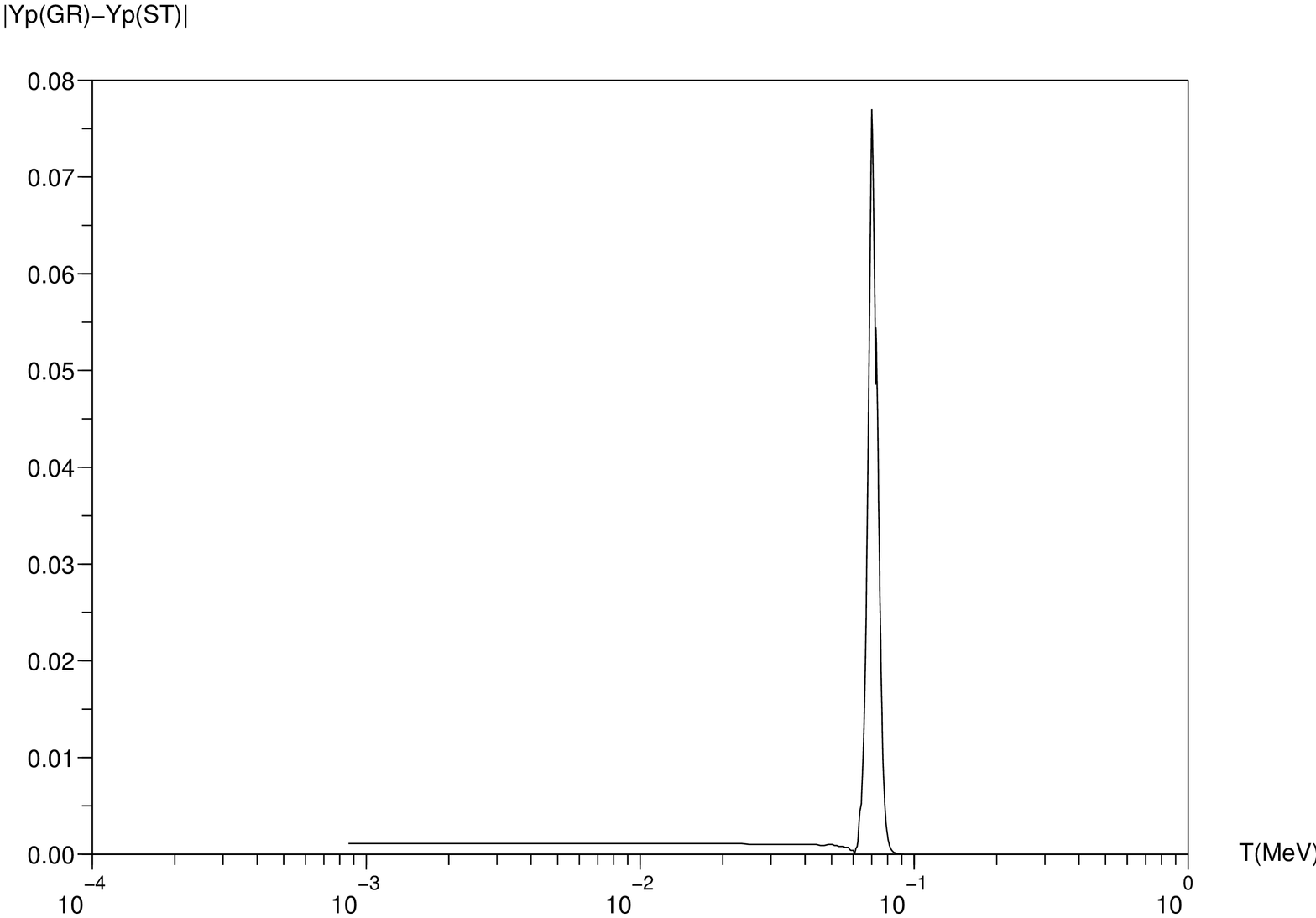}\\
\includegraphics[width=6cm,angle=270]{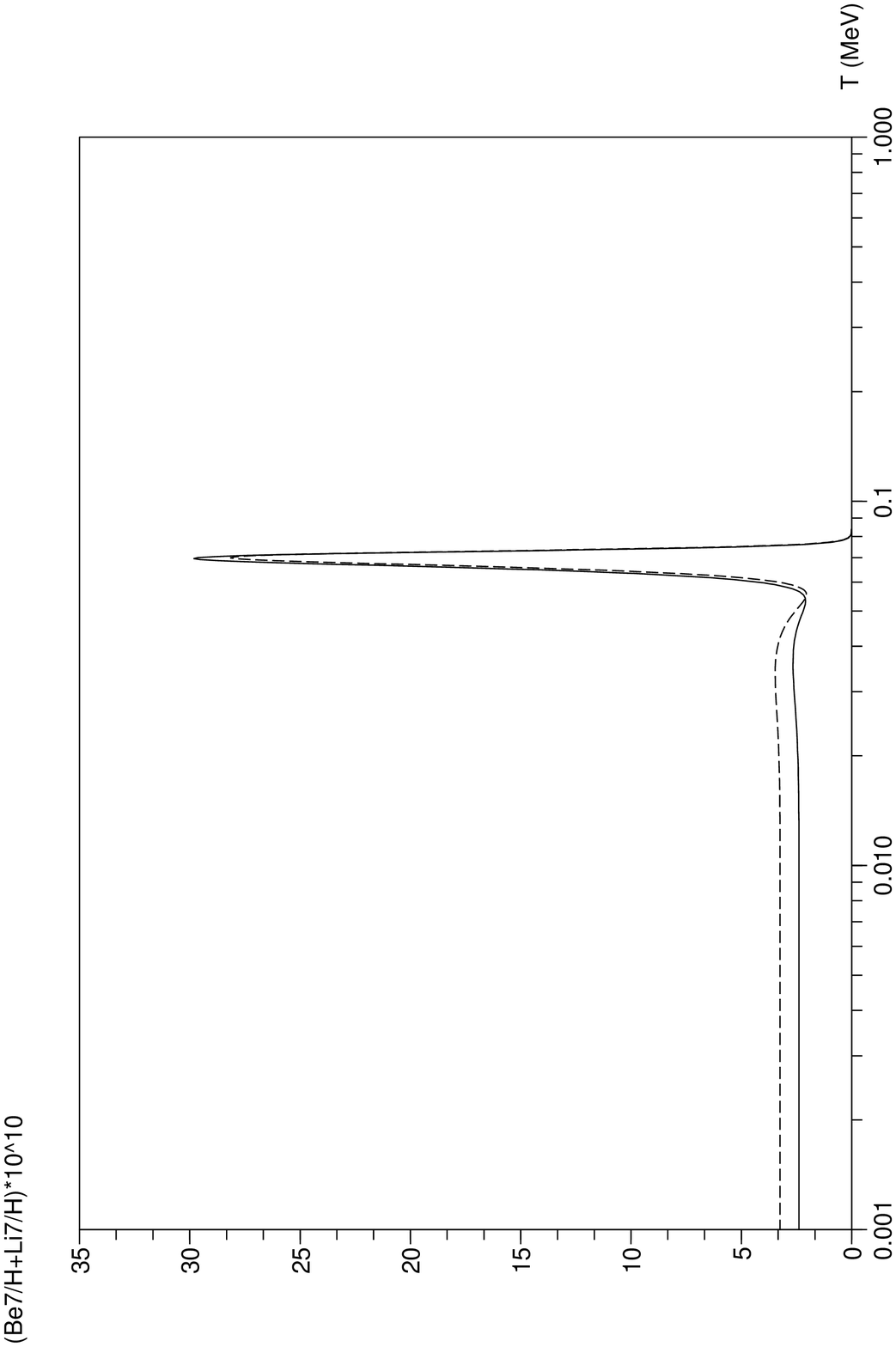}
\end{center}
\caption{Upper panel: Evolution of (n/p) as a function of the
temperature in General Relativity (dashed line) and in the same
model as in Figure \ref{fig:SUFEF} (solid line). The dashed dotted
line represents the (n/p) ratio at equilibrium. Medium panel:
Evolution of the difference between $^4$He abundances in General
Relativity and in the same model as in Figure \ref{fig:SUFEF}. Lower
panel: Evolution of the abundance of $^7$Be+$^7$Li as a function of
temperature in General Relativity (dashed line) and in the same
model as in Figure \ref{fig:SUFEF} (solid line). } \label{fig:np}
\end{figure}

The clues for the success of this kind of scalar-tensor theories in
solving the lithium problem have been already advanced at the
beginning of this subsection. The evolution of the ($n/p$) ratio, as
well as that of the $^4$He overproduction in the scalar-tensor model
exemplified here, are shown in figure (\ref{fig:np}). As can be noticed on
that figure, the effect of the speed-up factor on the n/p ratio is an
integrated effect at temperatures between a few MeV and a few hundreds of keV.
The fact that the speed-up factor is
less than 1 at temperatures higher than 1 MeV implies that the n/p ratio
departs from thermal equilibrium slighly later than in General Relativity, and
this effect is partly compensated by a speed-up factor greater than one after 1 MeV,
resulting in a n/p ratio that decreases faster in General Relativity than in the scalar-tensor
model between 1 MeV and 0.1 MeV. These two effects almost
compensate each other and result in a n/p ratio at the beginning of BBN comparable in the two models,
and so, in a $^4$He abundance in the scalar-tensor model that is not very
different from the one obtained in General Relativity. 
During the whole BBN period, the
speed-up factor is instead significantly higher than unity. Then,
the burning of $^4$He to produce heavier elements is less efficient
than in General Relativity, and the resulting final $^7$Li abundance
is low enough to be consistent with the observed value. Indeed, the
reaction $^3 \mbox{He}(\alpha,\gamma) ^7 \mbox{Be}$ dominates
the creation of $^7$Be (and then of $^7$Li by $\beta$-decay of $^7$Be) during
BBN. Therefore, neglecting the destruction of $^7$Be through the reaction
$^7 \mbox{Be}(\mbox{n,p})^7 \mbox{Li}$ that is inefficient because of the low
neutron abundance, a rough estimate of the final abundance of these
elements can be obtained from $dY_{Be7}/dT=\Gamma
Y_{He3}Y_{He4}/(TH)$ or, for small variations:

\begin{equation}\label{eq:dynabundances}
\Delta Y_{Be7} = \frac{\Gamma}{H}Y_{He3}Y_{He4}\Delta \ln T\
\end{equation}
where $\Gamma$ is the reaction rate of $^3 \mbox{He}(\alpha,\gamma)
^7\mbox{Be}$, while $Y_{Be7}$, $Y_{He3}$ and $Y_{He4}$ denote the
abundances of $^7$Be, $^3$He and $^4$He, respectively.

The lower panel of figure \ref{fig:np} shows the evolution of the
$^7$Be+$^7$Li abundance in this scalar-tensor model as well as in
General Relativity (GR). It can be seen from this figure that both
predictions start to significantly deviate only at temperatures
smaller than $0.06\mbox{ MeV}$, and reach their final constant
values at about $T\sim0.03\mbox{ MeV}$. By applying the equation
(\ref{eq:dynabundances}) within such a small range of temperatures,
we have:

\begin{equation}\label{eq:rapdeltaY}
\frac{\Delta Y_{Be}}{\Delta Y_{Be}^{GR}}\sim
\xi^{-1}\frac{Y_{He3}Y_{He4}}{Y_{He3}^{GR}Y_{He4}^{GR}},
\end{equation}
that also provides an estimate of the final $Y_{Li7}/Y_{Li7}^{GR}$
ratio. At temperatures below $0.06\mbox{ MeV}$, the abundances of
$^3$He and $^4$He are constant in both models, with $Y_{He3}\sim
Y_{He3}^{GR}$ and $Y_{He4}\sim Y_{He4}^{GR}$, while the speed-up
factor oscillates around $\xi\sim 1.16$. Therefore, equation
(\ref{eq:rapdeltaY}) implies $Y_{Li7}\sim 0.8 Y_{Li7}^{GR}$, in
rough agreement with our result obtained from the full integration
of the nuclear reaction network: $Y_{Li7}=0.74Y_{Li7}^{GR}$.

Of course, other theories with a different coupling function and/or
a different potential could have the same effect on the $^7$Li
abundance, provided that they lead to a speed-up factor with a time
evolution similar to that found in the models analyzed in this
section. As an example, the Appendix shows that, when the
$\dot{\varphi}_{init}=0$ condition is relaxed in scalar tensor
theories without a self-interaction term (then, the convergence
towards General Relativity is not ensured, and requires additional
conditions), some of the resulting models can also solve the lithium
problem. Therefore, the mechanism found here, based on a specific
shape of the variation of the expansion rate during BBN, can be
achieved in a great variety of scalar-tensor cosmological models.

\subsubsection{Beyond BBN processes: effects on CMB and matter power spectrum}

It is important to note that, in the analysis of the previous subsection, at
the end of BBN, $\alpha (\varphi)<10^{-5}$ and $A(\varphi)\sim 1$, so that
$G_{eff}/G_{N}\sim 1$; moreover, $\alpha (\varphi)$ goes on decreasing because
$\varphi$ decreases, so that one can consider that gravitation is indistinghishable from General Relativity already
at the end of BBN. Nevertheless, the scalar field still has a non negligible
energy density, and consequently, the expansion rate is not the one of General
Relativity with matter dust and radiation fluids, but receives an additional
contribution from the scalar field energy density. This is why the speed-up
factor is not immediately $1$ but oscillates around a value of $1.16$ until
the end of the radiation dominated era and converges towards 1 when dust
matter begins to dominate.
As a consequence, we expect that such a model will have some impact on
observables such as the matter power spectrum or the CMB. This is actually the
case, but, these impacts are really of a different nature that the ones that were previously
investigated in the context of scalar-tensor theories
\citep{Chen1999,Baccigalupi2000,Torres2002,Nagata2002,Nagata2003}: these works
address the effects of modification of gravity (mainly variations of the gravitational
coupling) on the CMB and matter power spectrum. In the models presented in
this paper, the variation of the gravitational coupling plays a very important
role during BBN (as examplified above), well before the matter dominated era,
but at the epoch of matter-radiation equality, the gravitational coupling no
longer significantly varies, and the scalar field behaves almost like a
standard scalar field ($\alpha(\varphi)\sim 0$).
Let's know consider the impact of our model on the matter power spectrum and
on the CMB. The matter power spectrum turn-over is defined by the scale
entering the Hubble horizon at the matter radiation equality (that is $H_{eq}^{-1}$)
\citep{Coble97}. Since the expansion rate is slightly faster in the model
presented here than in General Relativity at the time of equality, we expect a shift in the turn
over given by \citep{Baccigalupi2000}:
\begin{equation}
\label{TurnOverShift}
\frac{\delta k_{T}}{k_{T}}=-\left(\frac{\delta
  H^{-1}}{H^{-1}}\right)_{eq}\mbox{, }
\end{equation}   
Which yields, if we define the speed-up factor at equality by $\xi_{eq}$:
\begin{equation}
\label{TurnOverShift2}
\frac{\delta k_{T}}{k_{T}}=\xi_{eq}-1\mbox{.}
\end{equation} 
With $\xi_{eq}\sim 1.075$, one then has: $\delta k_{T}/k_T\sim7.5\%$. This
small shift is completely compatible with the current incertainties on data \citep{Tegmark2006}.

The position of the first acoustic peak in the CMB power spectrum is also
affected. The acoustic oscillations occur at an angular scale proportional to
the size of the CMB sound horizon at decoupling and inversely proportional to
the comoving distance between the observer and the last scattering
surface. So, if one defines the conformal time:
\begin{equation}
\label{Comovingtime}
\tau=\int_{0}^{t}\frac{dt}{a(t)}\mbox{, }
\end{equation} 
the multipole moment associated with the first peak is given by:
\begin{equation}
\label{lp}
l_{p}\propto \frac{\tau_{0}-\tau_{dec}}{\tau_{0}}
\end{equation}
where $\tau_{0}$ is the conformal time of the observer (today), and
$\tau_{dec}$ the conformal time at decoupling. Then the shift in the position
of the first peak: $\delta l_{p}/l_{p}=(l_{p}-l_{p}^{GR})/l_{p}^{GR}$ can be
numerically inferred from the simulation. For the model presented above, one
finds: $\delta l_{p}/l_{p}\sim 8\%$.  This is a rather important shift \citep{Page2003}, but it should be stressed that it depends strongly on the value of the expansion rate
today, and that it has been assumed, to make a significant comparison that this expansion rate today was the
same in the two models (General Relativistic cosmology and scalar-tensor
cosmology). It is possible to play with the acceptable value of $H_{0}$ (that
is to give different values of $H_{0}$ to the two models) to make this shift
smaller. 

Finally, one expects that the model presented above will lead to other
distortions in the CMB power spectrum at small scales since the expansion rate is
different from the one of General Relativity during the whole radiation
dominated era, when the small scales (those smaller than the one of the firsts
acoustic peak) are entering the sound horizon, but investigating these problem
demands a detailed analysis of the CMB physics in this new context that is
beyond the scope of this paper, and will be
dealt with in a forthcoming paper.

\section{Conclusion}

In the framework of the standard BBN, the WMAP measurement of the
baryon density of the Universe $\Omega _{b}h^{2}=0.0224\pm 0.0009$
leads to a serious discrepancy between the predicted and observed
$^7$Li abundance, even when systematic errors are cautiously taken
into account. We addressed this problem by renewing the standard BBN
scenario in a simple way: by considering that gravitation is
described by a scalar-tensor theory. The expansion of the Universe
during BBN is then modified, then modifying the conditions and the
efficiency of nuclear reactions while assuming the standard nuclear
physics. We showed that it is possible to obtain a small enough
$^7$Li abundance and, at the same time, to preserve the $^4$He and D
abundances thanks to a scalar-tensor theory of gravity with a non
monotonic speed-up factor that has a generic behavior. The expansion
must be similar to that found in General Relativity at
$T\sim1-2$MeV, so that the freezing-out temperature of the weak
processes $n+\nu_{e}\leftrightarrow p+e^{-}$ and
$n+e^{+}\leftrightarrow p+\bar{\nu}_{e}$ is also similar to that
obtained in General Relativity, then leading to almost the same
$^4$He production. However, during BBN, the expansion must be faster
than in General Relativity, making the reaction $^3
\mbox{He}(\alpha,\gamma) ^7 \mbox{Be}$ less efficient to burn $^4$He
and to produce $^7$Li.

Consequently, the $^7$Li abundance obtained for the baryon density
inferred from WMAP could be a good imprint for the presence of a
scalar field explicitly coupled to matter at the beginning of the
Universe, challenging our understanding of the primordial Universe
and the nature of gravity at early epochs. It could then be valuable
to study  the possible signatures of this scalar field in
inflationary scenarios and to perform a detailled analysis of its impact on
CMB physics and structure formation.

\section*{Aknowledgments}
We would like to thank Pr. G. Steigman for valuable comments on
this work.

\section*{APPENDIX: Varying speed-up factor in theories without a self-interaction term}

All the models analyzed in this paper starts from
$\dot{\varphi}_{init}=0$. As a consequence, in any theory without a
self-interaction term, the scalar field has a constant value during
the whole radiation-dominated period. Nevertheless, previous works
\citep{sernaalimi2,sernaalimi3} have shown that, when the condition
$\dot{\varphi}_{init}=0$ is relaxed, many of these theories also
imply a non-monotonic evolution of the speed-up factor. The
convergence of these theories towards General Relativity has instead
important differences with respect to that found for the models
analyzed in this paper. In non-monotonic scalar-tensor theories
without a self-interaction term, the convergence towards General
Relativity is not ensured independently of the initial conditions.

One of the main conclusions of this paper is that the lithium
problem can be solved in many scalar-tensor theories, defined by
different coupling functions and/or different potentials, but
implying an evolution of the speed-up factor with the same general
properties as those found in Section \ref{sec:selfinteracting}. In
this Appendix, we explore the impact on BBN of a theory defined, in
the Dicke-Jordan frame, by
\begin{equation}\label{eq:amodels}
3+2\omega(\phi)=\frac{a}{|\phi-1|}+b
\end{equation}
with $\phi_{init}>1$ (this is a necessary condition for the speed-up
factor to be less than unity at the beginning).

In the limit close to general relativity ($\phi\rightarrow 1$), the
above theory leads \citep{sernaalimi3bis,barrowparsons97} to an
Einstein frame formulation with $\alpha(\varphi)\propto
|\varphi|^{1/2}$ and $\varphi_{init}<0$. The effective potential
defined in (\ref{eq:effpot}) then becomes
\begin{equation}
V_{eff}(\varphi)=\mbox{sgn}(\varphi)\frac{2(1-3w_{b})}{3}|\varphi|^{\frac{3}{2}},
\end{equation}
where sgn$(x)=-1,0,1$  if $x$ is negative, zero or positive,
respectively. This effective potential does not have a minimum, but
a stationary point at $\varphi=0$. So, if the scalar field does not
have a high enough initial velocity, it will roll down its effective
potential during the matter-dominated era, pushing the theory far
from General Relativity. Then, in these theories, one must give an
initial velocity to the scalar field so that it can reach the
opposite side of the stationary point $0$ before the matter
dominated era. In \citet{damournordtvedt93}, it was shown that,
during the radiation dominated era, the total displacement of the
scalar fields is given by:
\begin{equation}\label{eq:deltaphi}
\Delta
\varphi=\frac{\sqrt{3}}{2}ln\left(\frac{1+\varphi_{i}^{'}/\sqrt{3}}{1-\varphi_{i}^{'}/\sqrt{3}}\right)
\end{equation}
So that we can deduce the initial velocity necessary to reach $0$,
starting from $\varphi_{i}$. It is given by:
\begin{equation}\label{eq:phidoti}
\varphi_{i}^{'}=\sqrt{3}\frac{exp(-2)\varphi_{i}/\sqrt{3}-1}{1+exp(-2)\varphi_{i}/\sqrt{3}}
\end{equation}
that is always finite and bounded by $-\sqrt{3}$ and $\sqrt{3}$.

Since it is not possible to perform the analytical transformation
(\ref{eq:conftrans}) to find its exact form in the Einstein frame,
the models defined by (\ref{eq:amodels}) have been integrated in the
Dicke-Jordan frame.  Moreover, the integration was performed
backward in time, then replacing the initial condition on the scalar
field by the present value of the coupling function $\omega(\phi)$.
The present value of the derivative of the scalar field was fixed to
zero.

\begin{figure}[h]
\begin{center}
\includegraphics[width=6cm]{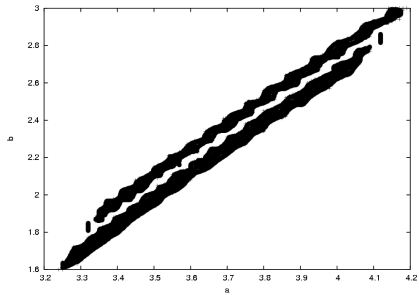}
\end{center}
\caption{Acceptable theories are represented by the shaded regions
in the $(a,b)$ plane, for  $3+2\omega(\phi)=\frac{a}{|\phi-1|}+b$ in
the Dicke-Jordan frame.} \label{fig:JFtheoplane}
\end{figure}

By varying the three remaining parameters $(a,b,\omega(\phi_{0}))$,
we identified the cosmological models that lead to the observed
abundances of light elements. The results were not sensitive to
$\omega(\phi_{0})$ as long as $\omega(\phi_{0})$ is sufficiently
large to pass the Solar System tests. Then, restricting the analysis
to the parameters $a$ and $b$, the figure (\ref{fig:JFtheoplane})
shows (as shaded regions) the $(a,b)$ couples leading to primordial
abundances of $^4$He, D and $^7$Li compatible with observations.
Again, the two distinct admissible regions are related to the two
different observations considered for the $^4$He abundance
\cite{Luridiana2003,Izotov1999}.

\vspace{3cm}

\begin{figure}[h]
\begin{center}
\includegraphics[width=8cm]{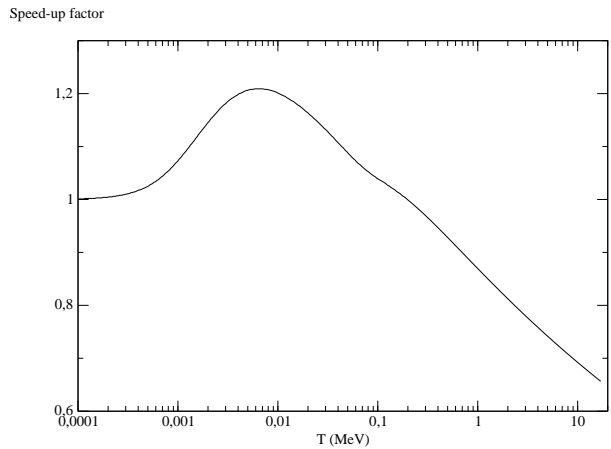}
\end{center}
\caption{Speed-up factor as a function of the temperature (in MeV)
during BBN.} \label{fig:SUFJF}
\end{figure}

As an illustration of these theories, the figure (\ref{fig:SUFJF})
shows the evolution of speed-up factor corresponding to the
particular choice $a=4$, $b=2.68$ and $\omega(\phi_{0})=10^{14}$. It
exhibits the behavior we described previously in order to solve the
lithium problem and to preserve the $^4$He abundance. The predicted
primordial abundances in this example were in fact:
$\mbox{Y}_{p}=0.2405$, $\mbox{D/H}=2.871\times 10^{-5}$ and
$^{7}\mbox{Li/H}=2.601\times 10^{-10}$.

\end{document}